\documentclass[twocolumn,showkeys,preprintnumbers,amsmath,amssymb]{revtex4}
\usepackage{graphicx}% Include figure files
\usepackage{dcolumn}% Align table columns on decimal point
\usepackage{bm}% bold math

\begin{document}

%\preprint{Lopac/1-2005}

\title{Chaotic dynamics of the elliptical stadium billiard \\ in the full
 parameter space}% Force line breaks with \\

\author{V. Lopac}
%\altaffiliation[Also at ]{Physics Department, XYZ University.}%Lines break
%automatically or can be forced with \\

\email{vlopac@marie.fkit.hr}
\affiliation{Department of Physics, Faculty of Chemical Engineering and
 Technology,\\ University of Zagreb, Croatia%
}%

\author{I. Mrkonji\' c}

\author{N. Pavin}%
\author{D. Radi\' c}%

%\homepage{}
\affiliation{
Department of Physics, Faculty of Science,\\ University of Zagreb,  Croatia
}%

\date{\today}% It is always \today, today,
	     
\begin{abstract} 
Dynamical properties of the elliptical stadium billiard, which is a
generalization of the stadium billiard and a special case of the recently
introduced mushroom billiards, are investigated analytically and
numerically. In dependence on two shape parameters $\delta$ and $\gamma$,
this system reveals a rich interplay of integrable, mixed and fully chaotic
behavior. Poincar\' e sections, the box counting method and the stability
analysis determine the structure of the parameter space and the borders
between regions with different behavior. Results confirm the existence of a
large fully chaotic region surrounding the straight line $\delta=1-\gamma$
corresponding to the Bunimovich circular stadium billiard. Bifurcations due
to the hour-glass and multidiamond orbits are described. For the quantal
elliptical stadium billiard, statistical properties of the level  spacing
fluctuations are examined and compared with classical results.    
\end{abstract}   
	     
%\pacs{05.45.-a; 03.65.-w}% PACS, the Physics and Astronomy
%Use
 %showkeys class option if keyword
\maketitle

%Section I
\section{Introduction}

An elliptical stadium is a two-parameter planar domain constructed by adding
symmetrically two half-ellipses to the opposite sides of a rectangle. In the
corresponding elliptical stadium billiard, the point particle is moving with
constant velocity within this boundary, exhibiting specular reflections  on
the walls. This billiard is a generalization of the Bunimovich stadium
billiard with circular arcs\cite{BuniSt}, which is fully ergodic for any
length of the central rectangle. When the semicircles are replaced by 
half-ellipses, a plethora of new dynamical properties emerges, ranging from
exact integrability, through mixed dynamics with a single chaotic component
or with a finely fragmented phase space, to broad regions of  fully
developed chaos. Thanks to these properties the elliptical stadium billiard
can be considered a  paradigmatic example of a Hamiltonian dynamical system,
which convincingly  illustrates both gradual and abrupt transitions induced
by the parameter variation.

To appreciate these advantages, the system should be looked at as a whole. 
Several descriptions of the elliptical stadium billiard can be found in the
literature \cite{Woj,Donnay,CanMarOKS,MarOKS,OKS,DelMaMar}. However,
although detailed and mathematically rigorous, they concentrate on some
specific billiard  properties and consider only restricted parameter
regions. Therefore in the present analysis we take into account all possible
boundary shapes and investigate the  properties of the elliptical stadium
billiard in the full parameter space spanned by two variables $\delta$ and
$\gamma$. Some of the preliminary results for special cases with the present
parametrization were described in \cite{LMRpar} and \cite{LMRg1}. The
elliptical  stadium billiard can also be considered as a special case of the
recently introduced mushroom billiards\cite{BuniMush,Lansel}.

Our interest for the elliptical stadium billiard has been enhanced by a
number of recent experiments and applications. Dynamical properties of
classical and quantal billiards have important consequences  for 
realistic systems in optics as well as in the atomic, mesoscopic and solid
state physics. Here we quote only a few examples. Details of the classical and
quantal billiard dynamics are the necessary tools for designing  properties
of semiconducting microlasers and optical devices used in communication 
technologies \cite{Gmachl,Hentschel}. Measured conductance fluctuations in
the semiconductor quantum dots of different shapes are obtained  from the
wave functions of the corresponding quantal billiards
\cite{Beenakker,Marcus}. The "atom-optics" billiards formed by laser beams
allow ultracold atoms to move within confined regions of desired shape
\cite{Milner,Friedman,Kaplan}. In the nonimaging optics, the billiard 
properties determine the shapes of the absolutely focusing
mirrors\cite{MirrBun}. As a consequence of the recent technological interest
in nanowires and similar extended microstructures, different problems with
the open billiards have also been  investigated\cite{Mendez,Kim}, including
the mushroom billiards \cite{Altmann}. Experiments performed with the
stadium-shaped microwave resonant cavities  \cite{Stock1,Stock2} could be
easily extended to stadia with elliptical arcs, predictably with interesting
results.

In Section II we define the billiard boundary and describe its geometrical
properties. In Section III the classical dynamics of the elliptical stadium
billiard  is presented. The existence and stability of some selected orbits
are investigated by means of the Poincar\' e plots and orbit diagrams.
Properties of the elliptic islands, their evolution and the bifurcations
induced by the variation of the shape parameters are presented. The extent
and limits of regions in the parameter space with different dynamical
behavior are shown. In Section IV further analysis of the Poincar\' e
sections is presented. The extent of the chaotic region of the phase space
is numerically estimated by means of the box-counting method.  In Section V
we consider the quantum-mechanical version of the elliptical stadium
billiard, present selected results for the level spacing statistics, and
compare them with the classical results for chaotic fraction of the phase
space. Finally, in Section VI we discuss the obtained results and propose
further  investigations.

%Section II
\section{Geometrical properties of the elliptical stadium}

The elliptical stadium is a closed planar domain, whose boundary in the $x-y$
plane is defined by means of two parameters $\delta$ and $\gamma$, satisfying
the  conditions $0\le\delta\le1$ and $0<\gamma<\infty$. It is symmetrical with
respect to the x- and y-axis and is described 
in our parametrization as

\begin{equation} 
y(x)=\left\{ \begin{array}{ll} \pm  \gamma &  {   \rm if\    } 
 0\le|x|<\delta \\ 
\pm\gamma\sqrt{1-{\left(\frac{|x|-\delta}{1-\delta}\right)}^2} &
 {   \rm  if\    }  \delta \le|x|\le1 \end{array} \right. 
\end{equation}

\noindent The meaning of parameters is visible from Fig. 1(a). The
horizontal diameter is normalized to 2. The vertical semiaxis of the ellipse
is  $\gamma$, and the height   $2\gamma$ of the billiard extends from $0$ to
$\infty$. The horizontal length of the central rectangle  is $2\delta$, and
the horizontal semiaxis of the ellipse is $1-\delta$. 

\begin{figure}
\caption{\label{fig:LopacFig1}(a) Shape parameters $\delta$ and
$\gamma$ for the elliptical stadium.
(b) Meaning of the angles $\phi$, $\phi$', $\theta$, $\alpha$ and
$\beta$. }
\end{figure}

The limiting  boundary subclasses  for some specific parameter combinations
are as follows. For $\delta=0$ the shape is elliptical, for $\delta=1$ it is
rectangular. Especially, for $\delta=\gamma=1$ it is a square, and for
$\delta=0$ and $\gamma=1$ a circle.  For $\delta=1-\gamma$ one obtains the
Bunimovich stadia\cite{BuniSt} with different lengths of the central 
rectangle, which define the border separating two distinct boundary classes,
one for $\delta<1-\gamma$ with elongated semiellipses, and the other  with
$\delta>1-\gamma$ and flattened semiellipses. Another interesting case is
$\delta=\gamma$, where the central rectangle is a square, and whose
properties were briefly discussed in \cite{LMRpar}.

The  focal points differ for the two classes. For $\delta>1-\gamma$ there
are four foci at the points

\begin{equation}
F\left[\pm\delta,\pm\sqrt{\gamma^2-(1-\delta)^2}\right] ,
\end{equation}

\noindent and for $\delta<1-\gamma$ the two  foci are situated at the points 

\begin{equation} 
F\left[\pm\left(\delta+\sqrt{(1-\delta)^2-\gamma^2}\right),0\right] .
\end{equation}

\noindent These expressions  enhance the importance of the term
$\tau=\gamma^2-(1-\delta)^2$ which is negative for $\delta<1-\gamma$, 
positive for $\delta>1-\gamma$, and zero for $\delta=1-\gamma$ (Bunimovich
stadium). For $\delta\le|x|\le1$, the  curvature radius is 

\begin{equation} 
R=\frac{[(1-\delta)^4+(|x|-\delta)^2[\gamma^2-(1-\delta)^2]]^{3/2}}
{\gamma(1-\delta)^4} .
\end{equation}

\noindent For $0\le|x|<\delta$ the boundary is flat and the curvature
radius is $R=\infty$.  At the endpoints of the horizontal and the vertical
axis of the ellipse the curvature radius is  $R_1=\gamma^2/(1-\delta)$ and 
$R_\delta=(1-\delta)^2/\gamma$, respectively. It is $R=1-\delta$ for the
circular arcs of the Bunimovich stadium. Fig. 2 shows some typical shapes of
the elliptical stadia for $\delta$ between 0 and 1  and for $\gamma$ between
0.25 and 1.5. The parameter $\gamma$ can have any value between  zero and
infinity. However, at $\gamma=0$ the boundary degenerates into a line, and
the values of $\gamma$ greater than 1.5 do not introduce any essentially new
features.

\begin{figure}
\caption{\label{fig:LopacFig2}Shapes of the elliptical stadium in
dependence on $\delta$ and  $\gamma$. Two of them ($\delta=\gamma=0.50$ 
and $\delta=0.75, \gamma=0.25$) are
Bunimovich stadia. }
\end{figure}

%Section III
\section{Classical dynamics of the elliptical stadium billiard}

Dynamics of a classical planar billiard can be examined by calculating  the
points on the billiard boundary where impacts and elastic reflections of the
particle take place. Fig. 1(b)  shows the meaning of variables describing an
impact and appearing in the conditions for existence and stability of
orbits. Symbols $\phi$ and $\phi'$, respectively, denote the angles which
the directions of the incoming and the  outcoming path make with the x-axis.
The normal to the boundary at the impact point T closes the angle
$\theta=(\phi+\phi')/2$ with the x-axis.  The angle between the normal and
the incoming (or outcoming) path is $\beta=(\phi'-\phi)/2$, and 
$\alpha=(\pi/2)-\beta$ is the angle between the tangent to the boundary at
the impact point T$(x,y)$ and the incoming (or outcoming) path.   The slope
of the normal to the boundary at the impact point is given as  the negative
inverse  derivative   $\tan\theta=-1/y'$. For the iterative numerical
computation of the impact points it is useful to know the relation between
the slopes of the incident and the outgoing path, expressed as 

\begin{equation} 
\frac{2\tan\theta}{1-\tan^2\theta}=\frac{\tan\phi+\tan\phi'}
{1-\tan\phi\tan\phi'} .
\end{equation}

\noindent In our numerical computations, two separate sets of data were
obtained: coordinates $x,y$ of the impact points  on the billiard boundary,
needed for the  graphical presentation of the orbits and for the computations
of the orbit stability, and coordinates  $X$ and $V_{\rm x}$, suitable for the
graphical presentation of the Poincar\' e sections.  The points P$(X,V_{\rm
x})$ in the Poincar\' e diagrams are obtained by plotting as $X$  the
x-coordinate of the point S in which the rectilinear path segment crosses  the
x-axis, and as $V_{\rm x}=\cos\phi$ the projection of the velocity on the
x-axis. With an additional assumption concerning  the horizontal diametral
orbit (explained in III.B) the Poincar\' e sections are thus completely
defined. As some orbit segments   do not cross the x-axis, the number of points
in the Poincar\' e diagram can be smaller than the number of the impact points
on the boundary. In our chosen system of units the mass and the velocity of the
particle  are $m=1$ and $V=1$, respectively, hence both $X$ and $V_{\rm x}$ lie
in the interval [-1,1]. For  billiards  with  noncircular boundary
segments\cite{LMRpar,LMRelh} such variables are computationally more convenient
than those containing the arc length variable suitable for the billiard
boundaries with circular arcs\cite{Berry}.  Since $X$  and $V_{\rm x}$ are 
canonically conjugated  variables and our billiard is a Hamiltonian system
which reduces to the collision-to-collision symplectic twist map, the phase
space and the corresponding Poincar\' e sections are area preserving.
 
Depending on the choice of the shape parameters and on initial conditions,
one obtains three types of orbits: periodic orbits, which define fixed
points in the Poincar\' e diagram; quasiperiodic orbits which define
invariant curves surrounding the fixed points, and chaotic orbits which fill
densely a part  of the phase space. As described in \cite{Berry}, periodic
orbits can be stable (elliptic), unstable (hyperbolic) and neutral
(parabolic). To discern these properties for the elliptical stadium
billiard, we use the criterion stated in \cite{Berry}, by which  the
stability  of an orbit is assured if the  absolute value of the  trace of
the deviation matrix $M$  is smaller than 2,  thus if

\begin{equation} 
-2<{\rm Tr} M<2 .
\end{equation}

\noindent The deviation matrix of the closed  orbit of period $n$ can be
written as  $M=M_{12}M_{23}..M_{n1}$, where the  $2\times2$ matrix $M_{ik}$
for  two subsequent impact points T$_i$ and T$_k$, connected by a
rectilinear path (the chord) of the length  $\rho_{ik}$, is \cite{Berry}

\begin{equation}
%\label{eq:}
M_{ik}=
\left(
\begin{array}{ll}
-\frac{\sin\alpha_i}{\sin\alpha_k}+\frac{\rho_{ik}}{R_i\sin\alpha_k}
 & -\frac{\rho_{ik}}{\sin\alpha_i\sin\alpha_k}\\
 & \\
-\frac{\rho_{ik}}{R_iR_k}+\frac{\sin\alpha_k}{R_i}+\frac{\sin\alpha_i}
{R_k} & 
-\frac{\sin\alpha_k}{\sin\alpha_i}+\frac{\rho_{ik}}{R_k\sin\alpha_i}
\end{array}\right) .
\end{equation}

\noindent To examine the properties of the elliptical stadium billiard in
the full parameter space, one may proceed by scanning the complete array of
$\delta$ and $\gamma$ values. However, it is often sufficient to make use of
a restricted, conveniently defined,  one-parameter subspace. Depending on
the  property considered, we choose among the following possibilities:
constant $\delta$ and changing $\gamma$, constant $\gamma$ and changing
$\delta$ as  in \cite{LMRg1}, or $\delta=\gamma$ as in \cite{LMRpar}.

\subsection{Billiards with $\delta<1-\gamma$}

This subfamily of elliptical stadia has elongated semiellipses. The
corresponding billiards  have been investigated in 
\cite{Donnay,MarOKS,CanMarOKS,OKS},  with the  principal aim  to establish
the limiting shapes beyond which the billiard is fully chaotic.  Parameters
$a$ and $h$ used there  to describe the boundary are connected with our
parameters $\delta$ and $\gamma$ through relations 

\begin{eqnarray}
\nonumber\\
\gamma=\frac{1}{h+a}, &&
\delta=\frac{h}{h+a} .
\end{eqnarray}

\begin{figure}
\caption{\label{fig:LopacFig3} Poincar\' e plots for $\delta<1-\gamma$,
obtained by plotting the pairs of variables $-1\le X \le 1$ and $-1\le V_{\rm
x} \le 1$ for $\delta=0.19$ and various values of  $\gamma$.  }
\end{figure}

The results reported in \cite{OKS,CanMarOKS} suggest that in the parameter
space there exists a lower limit above which the billiard  is chaotic, as 
consequence of the existence of the stable pantographic orbits. In Fig. 3
results of our numerical calculations of the   Poincar\' e sections are
shown for  $\delta=0.19$  and  different values of $\gamma<1-\delta$. This
example conveniently  illustrates  the behavior  typical for this region of
the parameter space. Numerous elliptic islands arising from the  stable
pantographic  orbits can be recognized. The islands due to  some other types
of orbits are also visible. Typical orbits contributing to these pictures
are shown in Fig. 4, for $\delta=0.19$ and increasing values of  $\gamma$.
The lowest two pantographic orbits, the bow-tie orbit  ($n=1$) and the candy
orbit ($n=2$), are shown in Figs. 4(f) and 4(h), respectively. Three higher
period pantographic orbits are seen in Figures 4(a), 4(c) and 4(d). Some
other types of orbits are depicted in Figs. 4(b), 4(e) and 4(g).  

\begin{figure}
\caption{\label{fig:LopacFig4} Some typical orbits appearing for
$\delta<1-\gamma$. Orbits in Figs. (a), (c), (d), (f) and (h) are 
pantographic orbits with $n=$ 4, 3, 2, 0 and 1, respectively. }
\end{figure}

\begin{figure}
\caption{\label{fig:LopacFig5}(a) Diagram of the two-dimensional parameter
space ($\gamma$, $\delta$). Thick lines denote the border of the fully
chaotic region including parts C, D and E. (b) Enlarged part of (a) showing
the emergence of multidiamond orbits of higher $n$. Diamonds, circles,
triangles, stars and crosses correspond to the cases illustrated on Figs. 3,
6, 8, 9, and 10,  respectively.}
\end{figure}

In \cite{MarOKS}, an earlier conjecture of Donnay\cite{Donnay}, stating that
the lower limit of chaos is set by relations $ 1<a<\sqrt{4-2\sqrt{2}}$ and
$h>2a^2\sqrt{a^2-1}$, was investigated. In our parametrization this region
is delimited by functions 

\begin{equation} 
\gamma=\frac{1-\delta}{\sqrt{4-2\sqrt{2}}} ,
\end{equation}

\begin{equation} 
\gamma=\frac{\sqrt{2}(1-\delta)^2}{\delta}\sqrt{\sqrt{1+\left(\frac{\delta}
{1-\delta}\right)^2}-1} ,
\end{equation}

\begin{equation} 
\gamma=1-\delta .
\end{equation}

\noindent In order to visualize the segmentation of the parameter space into
regions of different dynamical behavior, in Fig. 5(a) we plot the pairs of
shape parameters as points in the $\gamma$ - $\delta$ diagram. Possible
parameter choices are situated within an infinitely  long horizontal band of
height 1. The tilted line connecting the points (0,1) and (1,0) holds for
Bunimovich stadia. The points below and above this line, respectively, are
the elongated and the flattened elliptical stadia. The region delimited by
functions (9), (10) and (11) is visible as a narow quasi-triangular area
denoted by D in Fig. 5(a). The obtuse angle of  this quasi-triangle is
situated at the point $\gamma=0.48711$, $\delta=0.47276$.  In a recent
paper  Del Magno and Markarian\cite{DelMaMar} proved exactly that for this
region of the parameter space the elliptical stadium billiard is fully
chaotic (ergodic with K- and Bernoulli properties). However, there are
strong indications that  the lowest limit of chaoticity is not consistent
with this line. It can be  probably  identified with the limit derived in
\cite{CanMarOKS},  separating regions B and C in Fig. 5(a) and  resulting
from the onset of the stable pantographic orbits. Whereas the lower limit of
the region D consists of two parts, this new limit is made of an infinite
number of shorter segments, corresponding to all possible pantographic 
indices $n$\cite{OKS}.  Our numerical calculations based on  the
box-counting method  (described in the next section) confirm the chaoticity 
of the region C and identify the border between  regions B and C in the
parameter space as the lower limit of chaos.  

Another conspicuous feature of billiards with $\delta<1-\gamma$ is the
stickiness of certain orbits and consequent fragmentation of the chaotic
part of the phase space into two or more separate sections. This occurs
below  a certain limiting combination of $\delta$ and $\gamma$. The exact
shape of this limit is not obvious. It is probably connected with the
straight line $\delta=1-\gamma\sqrt{2}$, pointed out in \cite{Donnay},
separating in Fig. 5(a) the region A from the region B, which can be blamed
to the discontinuities in the boundary curvature at the  points where the
flat parts join the elliptical arcs. Similar  phenomena  have been  noticed
in various billiards with circular arcs\cite{DullRW,Reichl}.

\subsection{Billiards with $\delta>1-\gamma$}

This part of the parameter space, comprising the flat half-ellipses, had not
been investigated previously, except for the  brief analysis  in \cite{Woj},
later cited in \cite{Mark}. As an example of the typical behavior, the
Poincar\' e sections for $\delta=0.19$ and various $\gamma>1-\delta$ are
shown in Fig. 6. Here appear some new types of orbits and we examine their
existence and stability. These are  the diametral orbits (horizontal,
vertical and tilted) of period two, the  hour-glass orbit of period four, 
the diamond orbit of period four, as well as  the whole family of
multidiamond orbits. In our further description, we will refer to the impact
coordinates $x$ and $y$ in the first quadrant  (instead of  $|x|$ and
$|y|$), with no loss of generality for the obtained results,  due to the
symmetries of the boundary and of the considered orbits.

\begin{figure}
\caption{\label{fig:LopacFig6}Poincar\' e plots  for $\delta=0.19$ and various
 $\gamma$, for $\gamma>1-\delta$.}
\end{figure}

\subsubsection{Diametral horizontal and vertical two-bounce orbits}

The horizontal two-bounce orbit obviously exists for all combinations of
$\delta$ and $\gamma$.  However, the trace of the deviation
matrix\cite{Berry}  is  equal to 

\begin{equation} 
{\rm Tr}M=2\left [ 2\left ( \frac{\rho}{R}-1 \right )^2-1 \right ] ,
\end{equation}

\noindent where $\rho=2$ and $R=\gamma^2/(1-\delta)$, so that the stability
condition  (6) reads

\begin{equation}
\frac{\rho}{2R}<1 .
\end{equation}

\noindent Hence the two-bounce horizontal diametral orbit becomes  stable
for

\begin{eqnarray}
\nonumber\\
\delta>1-\gamma^2 ; &&
\gamma>\sqrt{1-\delta} .
\end{eqnarray}

\noindent Thus, there is a bifurcation at the value $\delta=1-\gamma^2$
giving birth to the stable diametral orbit. Again we choose $\delta=0.19$ as
a typical example, and keeping it constant we notice that for this value of
$\delta$ the upper limit of chaos is $\gamma=\sqrt{1-0.19}=0.9$. In the
region $1-\gamma<\delta<1-\gamma^2$, denoted in Fig. 5(a) by E, there are no
periodic orbits, and this is the region of full chaoticity.  This  result
has been proved in \cite{Woj}, among results for several chaotic billiards,
and later cited in a discussion of the elliptical stadium billiards
\cite{Mark}. (It should be mentioned that  for this case the  definition of
the shape parameter in \cite{Mark} differs from that in Eqs.(8).) 

In the Poincar\' e plot, the invariant points of the horizontal diametral
orbit are defined as $(0,\pm1)$, since this orbit can be understood as a
limiting case of  a long and thin horizontal hour-glass orbit, which will be
discussed below in more detail.  The Poincar\' e diagrams in Fig. 6 show
that the corresponding elliptic island arising for higher values  of
$\gamma$ above the $\gamma=\sqrt{1-\delta}$ line develops into a broad band filled
with invariant curves. The corresponding orbit will be shown below in Fig.
11(a). 
 
Here we mention also the existence of a family of the vertical two-bounce 
non-isolated neutral periodic orbits. Their properties are identical to
those of the bouncing-ball orbits  between the flat segments of the boundary
in the Bunimovich  stadium billiard\cite{Berry} and at present  do not
require our further attention.

\subsubsection{Tilted diametral two-bounce orbits}

A tilted two-bounce orbit  having the impact point T$(x,y)$  on the billiard
boundary with  derivative $y'$ exists if the following condition is fulfilled:

\begin{equation}
yy'+x=0 .
\end{equation}

This is satisfied if 

\begin{equation}
x=\frac{\gamma^2\delta}{\gamma^2-(1-\delta)^2}.
\end{equation}

\noindent This point should be on the elliptical part of the boundary 
$\delta<x<1$, which leads to the condition

\begin{equation} 
\gamma>\sqrt{1-\delta} .
\end{equation}  

\noindent However, from the stability condition (13),  where the chord
length is

\begin{equation}
\rho=2\gamma\sqrt{1+\frac{\delta^2}{\gamma^2-(1-\delta)^2}}  ,
\end{equation}

\noindent arises the restriction

\begin{equation}
\gamma<\sqrt{1-\delta},
\end{equation}

\noindent which cannot be fulfilled simultaneously with Eq.(17). The
conclusion is that no stable tilted two bounce orbit can exist.

\subsubsection{Diamond  orbit}

The diamond orbit of period four shown in Fig. 7(a) exists for any parameter
choice. It has two bouncing points at the  ends of the horizontal semiaxis,
and the other two on the flat parts on the boundary.  To assess its
stability, one should calculate the deviation matrix $M=(M_{01}M_{10})^2$.
The angles contained in the matrix are given as $\sin\alpha_0=\gamma/\rho$
and $\sin\alpha_1=1/\rho$, where $\rho=\sqrt{1+\gamma^2}$. The curvature
radius at the point  $x=1$ is $R=\gamma^2/(1-\delta)$. This leads to the
trace of the matrix

\begin{figure}
\caption{\label{fig:LopacFig7} Multidiamond orbits with $n=$   1 - 8.}
\end{figure}

\begin{equation} 
{\rm Tr}M=2\left [ 2\left ( \frac{2\rho^2}{R}-1 \right )^2-1 \right ] 
\end{equation}

\noindent and to the condition $\rho^2<R$ or $1+\gamma^2<\gamma^2/(1-\delta)$,
 thus the  diamond orbit becomes stable when

\begin{eqnarray}
\nonumber\\
\delta>\frac{1}{1+\gamma^2}; &&
\gamma>\sqrt{\frac{1}{\delta}-1} .
\end{eqnarray}

\noindent This limiting curve is shown in Fig. 5(a) as a full line dividing
the region F from G and the region H from I. 

\subsubsection{Multidiamond orbits}

The multidiamond orbit of order $n$ is the orbit of period $2+2n$,  which
has two bouncing points at the ends of the horizontal axis and $2n$
bouncing points on the flat parts of the boundary (Fig. 7). Such orbit
exists if

\begin{equation}
\delta>1-\frac{1}{n}.
\end{equation} 
 
\noindent Thanks to the trick known from  geometrical optics\cite{BuniMush}
which  allows the mirroring of the billiard around the flat part of the
boundary, the calculation of the deviation matrix for the order $n$ becomes
identical to the calculation of the matrix for the diamond orbit. Thereby
the chord  length $\rho$ in (20) should be replaced by

\begin{equation} 
L=n\rho_n ,
\end{equation}

\noindent where

\begin{equation}
\rho_n=\sqrt{\frac{1}{n^2}+\gamma^2}.
\end{equation}    

\noindent The trace of the deviation matrix is then 

\begin{equation} 
{\rm Tr}M=2\left [ 2\left ( \frac{2\rho_n^2n^2}{R}-1 \right )^2-1 \right ] 
\end{equation}

\noindent with $R=\gamma^2/(1-\delta)$. The resulting condition for the
stability of the multidiamond orbit is then 

\begin{eqnarray}
\nonumber\\
\delta>1-\frac{\gamma^2}{1+\gamma^2n^2} ; &&
\gamma>\sqrt{\frac{1-\delta}{1-n^2(1-\delta)}}.
\end{eqnarray}

\noindent We stress again that the special case $n=1$ corresponds to the
diamond orbit. The limiting curves (26) are plotted in Fig. 5(a) and are
shown enlarged in Fig. 5(b). For $\gamma\to\infty$ the minimal values of  
$\delta$ after which the multidiamond orbits appear are

\begin{equation} 
\lim \limits_{\gamma \to\infty}\delta=1-\frac{1}{n^2} .
\end{equation}

\noindent The emergence of the stable multidiamond orbits can be followed by
observing the Poincar\' e sections for  a set of values $\delta=\gamma$
(Fig. 8). The values of $\delta$ for which an orbit of new $n$ appears are
given as intersections of the straight line $\delta=\gamma$  with curves
defined by (26), and satisfy the equation

\begin{figure} 
\caption{\label{fig:LopacFig8}Poincar\' e plots for a set of different
values $\delta=\gamma$,  showing the appearance of successive stable
multidiamond  orbits of higher order.}
\end{figure}

\begin{equation} 
n^2\delta^3-(n^2-1)\delta^2+\delta-1=0 .
\end{equation}

\noindent Especially, for the diamond  orbit ($n=1$) this equation reads

\begin{equation} 
\delta^3+\delta-1=0 
\end{equation}

\noindent and the orbit becomes stable for
$\delta=\gamma=(u-1/u)/\sqrt{3}=0.6823278$, where
$2u^3=\sqrt{31}+\sqrt{27}$. For the same type of boundary the stable
two-bounce orbit appeared at $\delta=\gamma=(\sqrt{5}-1)/2=0.618034$.

\subsubsection{The hour-glass orbit}

The hour-glass orbit  looks like the bow-tie orbit rotated by $\pi/2$. It 
exists if the coordinates $x$ and $y$ of the impact point on the boundary
and the derivative $y'$ of the boundary at this point satisfy the equation

\begin{equation}  
2xy'+yy'^2-y=0 ,
\end{equation}

\noindent giving as solution the x-coordinate of the point of impact

\begin{equation} 
x=\delta+\frac{(1-\delta)^2}{\gamma^2-(1-\delta)^2}\left 
(\delta+\sqrt{\delta^2+[\gamma^2-(1-\delta)^2]}
\right ) .
\end{equation}

\noindent The condition $\delta<x<1$ that this point should lie on the
elliptical part of the boundary leads to the requirement 

\begin{eqnarray}
\nonumber\\
\delta>1-\frac{\gamma^2}{2}; &&
\gamma>\sqrt{2(1-\delta)} .
\end{eqnarray}

\noindent This limit is shown in Fig. 5(a) as a dashed line separating 
 the region G from I 
and region F from H. If we denote the points with positive $y$ by
1 and the points on the negative side by -1, the deviation matrix can be
calculated as 

\begin{equation} 
M=(M_{11}M_{1-1})^2 
\end{equation} 

\noindent with $x$ given by Eq. (31). The corresponding angle $\alpha$
needed in the matrix (7) is  given by

\begin{equation} 
\sin\alpha=\frac{\gamma(x-\delta)}{\sqrt{(1-\delta)^4+
(x-\delta)^2[\gamma^2-(1-\delta)^2]}}.
\end{equation} 

\noindent The chords lengths are
 
\begin{equation}  
\rho=\rho_{1,1}=2x 
\end{equation} 

\noindent and 

\begin{equation}  
\rho'=\rho_{1,-1}=2\sqrt{x^2+y^2} ,
\end{equation}

\noindent where $x$ is given by (31) and $y$ is calculated from (1). The
curvature radius at this point is  obtained by substituting (31) into (4).
If we define 

\begin{equation}  
\Phi=\frac{\rho}{R\sin\alpha}\frac{\rho'}{R\sin\alpha}-(\frac{\rho}
{R\sin\alpha}+
\frac{\rho'}{R\sin\alpha})
\end{equation}

\noindent the trace of the deviation matrix is 

\begin{equation}  
{\rm Tr}M=2\left [ 2(2\Phi+1)^2-1  \right ]  .
\end{equation} 

\noindent Whereas the left hand side of the stability condition (6) is valid
automatically, its right hand side is fulfilled only if 

\begin {equation}
-1<\Phi<0 .
\end{equation}

\noindent The left hand side limit of (39) $\Phi=-1$   reduces to the
existence condition (32) and  denotes the case where the hour-glass orbit 
degenerates into the horizontal diametral orbit. The right hand side limit 
$\Phi=0$   is identical to the condition 

\begin{equation} 
\frac{\rho}{R\sin\alpha}\frac{\rho'}{R\sin\alpha}=\frac{\rho}{R\sin\alpha}+
\frac{\rho'}{R\sin\alpha} ,
\end{equation} 

\begin{figure}
\caption{\label{fig:LopacFig9} Enlarged parts of the Poincar\' e plots,
 showing the stable island due to the
 hour-glass orbit, for $\delta=0.19$ and $1.30<\gamma<1.80$.}
\end{figure}

\begin{figure} 
\caption{\label{fig:LopacFig10} Highly enlarged parts of the Poincar\' e
plots, showing the stable  island due to the hour-glass orbit, for
$\delta=0.19$ and $1.87<\gamma<2.00$.} 
\end{figure}

\noindent giving the prescription for the numerical evaluation of the
limit  beyond  which the hour-glass  orbit becomes unstable. This is
illustrated on Figs. 9 and 10, where Poincar\' e sections are shown for
typical examples with  $\delta=0.19$ and a set of boundary shapes  between
$\gamma=\sqrt{2(1-\delta)}=1.2728$ (where the stable orbit emerges) and
$\gamma=1.993$ (where it becomes hyperbolic). The island corresponding to
the hour-glass orbit is shown enlarged, with different scales in Fig. 9
and Fig. 10. The evolution of the hour glass orbit has very interesting
properties. At first this island appears as a tiny point at the upper end
of the Poincar\' e diagram, then descends following the value $V_{\rm
x}=\cos\phi$ of its vertical coordinate. When  $\gamma$ is varied, the
central island acquires a resonant belt, with periodicity which changes 
from 8 to 7, then to 6, 5 and 4, as shown in Fig. 9. Near $\gamma=1.79$  
the orientation of the rectangular island changes from tilted to
horizontal, after which a triangular island appears. Near $\gamma=1.90$
(Fig. 10) the triangular island shrinks to infinitely small size, and then
starts growing again, but with the opposite  orientation. This is the
phenomenon of the "blinking island" already noticed in \cite{LMRelh} and
described in \cite{ZasEN}.  With further increase of $\gamma$ an
interesting type of bifurcation takes place. The typical islands in the
Poincar\' e  sections corresponding to these changes are visible in Fig.
10, and the orbits in Fig. 11. At the value $\gamma=1.993$, consistent with
the upper stability limit expressed by (40), the stable hour-glass orbit in
Fig. 11(e) is replaced by an unstable hour-glass orbit, shown in Fig. 11(f)
for $\gamma=2.02$. Simultaneously with this, two stable orbits  appear.
They have lost the symmetry of the hour-glass orbit, but in the mutual
relation to each other retain the left-right mirroring  symmetry (Figs.
11(g) and 11(h)). This type of behavior, known as the pitchfork
bifurcation, has been  encountered in  lima\c con billiards\cite{Baecker}.

\begin{figure}  
\caption{\label{fig:LopacFig11}(a) Stable orbit  into which degenerate both
the two-bounce horizontal orbit and  the hour-glass orbit; (b) The
hour-glass orbit near the lower limit of stability;  (c),(d) Orbits due to
the resonant belts shown in Fig. 9; (e) Stable hour-glass orbit at the upper
stability limit; (f) Unstable hour-glass orbit beyond the upper stability
limit; (g),(h) Two sligthly deformed  mutually symmetrical orbits, formed in
the bifurcation of the hour-glass orbit.}
\end{figure} 

%Section IV
\section{Analysis of  the parameter space properties by means of 
the box counting method}

In this section we return to the question of limits within which the
elliptical stadium billiard is fully chaotic.  For $\delta>1-\gamma$,   our
numerical calculations of the billiard dynamics  confirmed the result of
\cite{Woj,Mark}, stating that the billiard is chaotic for
$1-\gamma<\delta<1-\gamma^2$. For $\delta<1-\gamma$, it was proved in
\cite{DelMaMar}  that the billiard is fully chaotic for the narrow region
determined by (9), (10) and (11) below the Bunimovich line $\delta=1-\gamma$.
In \cite{CanMarOKS,OKS}  this band was conjectured to be  much broader, and
its lower limit attributed to the emergence of the stable pantographic
orbits. 

Here we propose to test these limits numerically, by means of the
box-counting method. We calculate the Poincar\' e sections for a chosen pair
of  shape parameters, starting with  5000 randomly chosen sets of initial
conditions and iterating each orbit  for 100 intersections with the x-axis,
thus obtaining 500000 points ($X$,$V_{\rm x}$) in the Poincar\' e diagram.
Having in mind the symmetry of the billiard, we plot the absolute values 
($|X|$, $|V_{\rm x}|$) of the computed pairs. Then we divide the obtained
diagram into a grid of small squares of side 0.01,  count the number $n$  of
squares which have points in them and calculate the ratio of this number to
the total number $N$ of squares.  The obtained ratio $n/N$ is denoted by
$q_{\rm class}$ and  determines the fraction of the chaotic region in the
Poincar\' e section. In this way also certain points  belonging to invariant
curves within the regular islands are  included. However, their  contribution
is negligible in comparison with the considerably  larger chaotic
contribution, since  the average  number of  points per square is 50 and our
principal aim is to examine the onset of complete chaos. Numerically, the
size of the box has been varied until the saturation of $n/N$ ratio was
achieved, providing a  consistent and reliable   method of  analysis of the
set obtained in the numerical experiment. We proceed by calculating  the
chaotic fraction for a constant value of $\delta$ and varying $\gamma$.
Results are shown in Fig. 12, where the classical chaotic fraction $q_{\rm
class}$  is plotted against $\gamma$ for a set of values of $\delta$.  These
pictures show that for each $\delta$ there exist  pronounced lower and upper
limits of the fully chaotic region, characterized by $q_{\rm class}=1$.  When
one plots these limits  in the parameter space diagram in Fig. 5(a), one
obtains two curves, the upper one  within the region  $\delta>1-\gamma$,
which coincides with  Eq. (14) and separates  the region E from region F, and
the lower one in the region  $\delta<1-\gamma$, separating regions B and C.
Thus our numerical experiments   confirmed the existence of both limits, the
lower one  determined in \cite{CanMarOKS}  from the pantographic orbits, and
the upper one suggested in \cite{Woj}. Outside this fully chaotic region
(consisting of regions C, D and E in Fig. 5(a)), the value of the fraction 
$q_{\rm class}$ is characterized by oscillations following the parameter
variation, called in \cite{DullRW} the "breathing chaos". At some values
there are strong discontinuities, which can be traced to  bifurcations of
various orbits. 

\begin{figure}
\caption{\label{fig:LopacFig12} Dependence of $q_{\rm class}$ on the shape
parameter $\gamma$ for various values of $\delta$,  calculated by means of
the box-counting method. There is a conspicuous difference between the cases
$\delta=0$ and $\delta=0.01$, typical for stadium billiards. For each
$\delta\ne0$ there exists an interval of full chaos, where $q_{\rm
class}=1$. }
\end{figure}  

%Section V
\section{The quantum-mechanical elliptical stadium billiard}

The elliptical stadium billiard can also be considered as a two-dimensional
quantal system.  The Schr\" odinger equation for a particle of mass $m$
moving freely within the two-dimensional billiard  boundary  is identical to
the Helmholtz equation for the wave function $\Psi$

\begin{equation} 
-\frac{\hbar^2}{2m}\nabla ^2\Psi = E \Psi, 
\end{equation}

\noindent where $E$ is the  particle energy.  The usual  transformation to
dimensionless variables is equivalent to substituting

\begin{equation}
\frac{\hbar ^2}{2m}=1\:\:\: {\rm and} \:\:\:   E=k^2 ,
\end{equation}

\noindent which yields

\begin{equation}
\left[ \frac{\partial^2}{\partial x^2}+\frac{\partial^2}
{\partial y^2}+k^2\right]\Psi(x,y)=0 .
\end{equation}

\noindent We solve this equation following the method of Ridell
\cite{Ridell} and, using the expansion in the spherical Bessel functions
\cite{LMRparlim}, obtain the wave functions and energy levels within the
preselected energy interval. The described method  yields about 500-3000
levels, depending on the shape.

\subsection{Energy level statistics for the elliptical stadium billiard}

According to \cite{BGS, Cas},  statistical properties of the energy spectrum
reflect the degree of chaoticity in the corresponding classical system. For
the elliptical stadium billiard it would be interesting to  see whether  the
quantal calculation confirms the existence  of the region of fully
developed chaos, in the sense of the conjecture of Bohigas, Giannoni and
Schmit\cite{BGS}.To assess this correspondence, we calculate  the
spectrum  and wave functions for a given pair of shape parameters. The
obtained spectrum is then unfolded using the method of French and
Wong\cite{FW} and the resulting histograms are analyzed by means of the
Nearest Neighbor Spacing Distribution Method (NNSD). To fit the numerically
calculated distributions, we propose three possibilities: the Brody
distribution\cite{Brody}, the Berry-Robnik distribution \cite{BeRo84}, and a
two-parameter generalization of both Brody and Berry-Robnik distributions

\begin{eqnarray}
P^{\rm PR}(s)={\rm e}^{-(1-q)s} (1-q)^2 Q \left[ \frac{1}{\omega+1},
 \alpha q^{\omega+1}s^{\omega+1}\right]+ \nonumber\\
 \nonumber\\
+{\rm e}^{-(1-q)s}q \left[ 2(1-q)+\alpha(\omega+1)q^{\omega+1}s^{\omega} 
\right] {\rm e}^{-\alpha q^{\omega+1}s^{\omega+1}}
\end{eqnarray}

\noindent which we shall henceforth call the Prosen-Robnik (PR)
distribution. For $\omega=1$, $P^{\rm PR}(s)$ reduces to Berry-Robnik
distribution, and for $q=1$ it is identical to Brody distribution.  It  
coincides with Wigner distribution if $\omega=1$ and $q=1$, and with the
Poisson distribution whenever $\omega=0$. Here, $\alpha $ is defined as  

\begin{equation} 
\alpha=\left[\Gamma\left(\frac{\omega+2}{\omega+1} \right)\right]^{\omega+1} 
\end{equation}

\noindent and $Q$ denotes the Incomplete Gamma Function

\begin{equation} 
Q(a,x)=\frac{1}{\Gamma(a)} \int_x^\infty{{\rm e}^{-t}t^{a-1}{\rm d}t} .
\end{equation}

\noindent The derivation of Eq.(44) was based on the factorized gap
distribution $Z^{\rm PR}(s)$

\begin{equation}
Z^{\rm PR}(s)={\rm e}^{-(1-q)s} Q \left[ \frac{1}{\omega+1},
\alpha q^{\omega+1}s^{\omega+1}\right]
\end{equation}

\noindent introduced by Prosen and Robnik\cite{ProRo93}. The gap
distribution is the probability that no level spacing is present in the
interval between $s$ and $s+\Delta s$. Relation between the level spacing
distribution $P(s)$ and the gap distribution is $P(s)={\rm d}^2Z(s)/{\rm
d}s^2$. The function (44) has been  evaluated in \cite{LBP96} and was used
to test the spectra of several types of the billiards of mixed
type\cite{LMRparlim,LMRelh}. Fitting the calculated histograms  to this
distribution gives two  parameters $q_{\rm PR}$ and  $\omega_{\rm PR}$.
According to \cite{ProRo93},  the resulting value of $q_{\rm PR}$  is the
variable which corresponds to the classical  $q_{\rm class}$,  the magnitude
of the chaotic fraction of the phase space. This distribution gives more
realistic results than the Brody  or the Berry-Robnik distribution.

The Berry-Robnik and Brody distributions have different behavior for  very
small spacings: Brody distribution vanishes at $s=0$ , whereas for the
Berry-Robnik distribution  $P^{\rm BR}(0)=1-q^2$.  Besides, the
Berry-Robnik  parameter $q$ has a well defined physical meaning:
quantitatively it is the fraction of the phase space which is filled with
chaotic trajectories, whereas the remaining regular fraction of the phase
space is equal to $1-q$.  However, the Berry-Robnik distribution is exactly
applicable only in the semiclassical limit, and we are exploring the
complete spectrum, including the lowest lying levels.

In \cite{ProRo93} Prosen and Robnik argue that distribution (44) describes
simultaneously transition from semiclassical to quantal regime and 
transition from integrability to chaos. The two parameters  $\omega$ and 
$q$ characterize these two transitions, respectively, so that $q$ retains
its meaning as the chaotic fraction of the phase space, but is applicable
also to cases far from the semiclassical limit.   The diagram of (44) in
dependence on $\omega$ and $q$ has been presented in \cite{LBP96}.

In applying the generalized distribution (44) to the billiard (1)  we hold
both parameters $\omega$ and $q$   within the limits [0,1]. We choose the
subset of elliptical stadium billiards with $\delta=\gamma$, and in Figs.
13(a)-13(k)  we show the histograms and their  fit with $P^{\rm PR}$. In
the region where $\delta=\gamma$ lies between 0.44 and 0.63 the distribution
is very close to Wigner and $q_{\rm PR}=1$. According to \cite{BGS} this
would correspond to the complete classical chaos. In Fig. 13(l) some
obtained values of $q_{\rm PR}$ are shown in dependence on $\delta$ and
compared with the classical result obtained with the box-counting method.
The values obtained with the Brody and Berry-Robnik distributions are also
shown. 

\begin{figure}
\caption{\label{fig:LopacFig13} (a)-(k) Histograms of the level density
fluctuations  for spectra of billiards with $\delta=\gamma$, fitted to the
Prosen-Robnik distribution (44); (l) Comparison of the classical  chaotic
fraction $q_{\rm class}$ with the values obtained by fitting the histograms
with Prosen-Robnik, Berry-Robnik and Brody distributions.}
\end{figure}

To the  question  whether  the fluctuations of $q_{\rm class}$  with 
gradual changes in shape parameter reflect on the quantal values of $q$
outside the chaotic region, the answer is positive, but requires further
investigation. Considering the importance of bifurcations  examined in
Section III, one can expect the effects similar to those found in the oval
billiard\cite{Makino} .

%Section VI
\section{Discussion and conclusions}

In this work we have explored classical and quantal dynamics of  the elliptical
stadium billiards in the full two-parameter space, analyzing two distinct
groups of these billiards, separated from each other by the set of Bunimovich 
billiards.  For  $\delta<1-\gamma$ we have confirmed the important role of the
pantographic orbits in establishing the lower bound for chaos. For
$\delta>1-\gamma$ we have analyzed in detail the diametral two-bounce orbits
and the diamond and multidiamond orbits of increasing periodicity, creating 
numerous new corresponding elliptic islands when  the value $\delta=1$ is
approached.  Especially interesting is the behavior of the hour-glass orbit
which is, while remaining stable within a large range of increasing values of 
$\gamma$, accompanied with a resonant belt with changing periodicity. At the
value where the orbit becomes unstable, a bifurcation to a pair of  stable
quasi-hour-glass orbits with distorted symmetry appear.  The upper limit of
chaos $\delta=1-\gamma^2$ is obtained by estimating the stability of the
horizontal two-bounce orbit, and coincides with the limit proposed in
\cite{Woj}. We have thus  numerically confirmed the existence   of a broad
region of chaos in the parameter space surrounding the straight line belonging
to the Bunimovich stadia.

Thereby the most important dynamical phenomena of the elliptical stadium
billiards in the full parameter space are revealed. There remains, however,
the open question of the stickiness of  orbits in the region
$\delta<1-\gamma$, the fragmentation of the phase space and its connection
with the discontinuities of the boundary curvature, and possible importance
of the limiting straight line  $\delta=1-\gamma\sqrt{2}$. One possible
method to resolve this  is to analyze the statistics and the phase space
properties of the leaking billiards \cite{Armstead, Tel}. Preliminary
investigations  show  promising results, and should be able to  explain the
foliation and the fragmentation of the phase space of the elliptical 
stadium billiard. 

In conclusion, our investigation has revealed dynamical properties of a
large two-parameter family of stadium-like billiards exhibiting a rich
variety of integrable, mixed and chaotic behavior, in dependence on two
shape parameters $\delta$ and $\gamma$, with the special case of
$\delta=1-\gamma$ corresponding to the Bunimovich stadium billiard 
\cite{BuniSt}. The proposed billiard shapes and obtained results could serve
as an additional testing ground for the experimental properties of
semiconducting optical devices and microwave resonant cavities.

%Section VII
\section{Acknowledgments}

We are grateful to A. B\" acker, V. Danani\' c, M. Hentschel, T. Prosen and
M. Robnik for enlightening discussions.  

%\newpage

\newpage

\end{document}